\documentclass[twocolumn,showpacs,floats,floatfix,superscriptaddress,aps,pra]{revtex4-1}
\usepackage{amsfonts}
\usepackage{amssymb}
\usepackage{amsmath}
\usepackage{calc}
\usepackage{graphicx}
\usepackage{bm}

\usepackage[normalem]{ulem}
\usepackage{amsmath,amssymb}
\usepackage{multirow}
\usepackage{xcolor,soul}
\usepackage{srcltx}
\usepackage{hyperref,graphicx}

\def\be{ \begin{equation}}
\def\ee{ \end{equation}}
\def\bea{ \begin{eqnarray}}
\def\eea{ \end{eqnarray}}
\def\bse{ \begin{subequations}}
\def\ese{ \end{subequations}}
\def\bc{ \begin{center}}
\def\ec{ \end{center}}

\begin{document}

\author{Stefano Longhi$^{*}$} 
\affiliation{Dipartimento di Fisica, Politecnico di Milano and Istituto di Fotonica e Nanotecnologie del Consiglio Nazionale delle Ricerche, Piazza L. da Vinci 32, I-20133 Milano, Italy}
\email{stefano.longhi@polimi.it}

\title{Kramers-Kronig potentials for the discrete Schr\"odinger equation}
  \normalsize


%
\bigskip
\begin{abstract}
\noindent  
In a seminal work, S.A.R. Horsley and collaborators [S.A.R. Horsley {\em et al.}, Nature Photon. {\bf 9}, 436 (2015)] have shown that, in the framework of non-Hermitian extensions of the Schr\"odinger and Helmholtz equations, a localized complex scattering potential with spatial distributions of the real and imaginary parts related to one another by the spatial Kramers-Kronig relations are reflectionless and even invisible under certain conditions. Here we consider the scattering properties of  Kramers-Kronig potentials for the discrete version of the Schr\"odinger equation, which generally describes wave transport on a lattice. Contrary to the continuous Schr\"odinger equation, on a lattice a stationary Kramers-Kronig potential is reflective. However, it is shown that a slow drift can make the potential invisible under certain conditions.
\end{abstract}



\maketitle

\section{Introduction}
Reflection is an ubiquitous phenomenon of wave physics which is found both in classical and quantum systems \cite{r1}. Reflection of electromagnetic (optical) waves in dielectric media with sharp refractive index changes and scattering of non-relativistic particles from a quantum potential provide important examples of wave reflection which share strong similarities \cite{r1,r2,r3,r4,r5}. However, it is known since long time \cite{r2} that reflection can be avoided in special classes of scattering potentials, the so-called reflectionless potentials \cite{r6,r7,r8,r9}. Recently, wave reflection and scattering from complex potentials in non-Hermitian systems has sparked a great interest with the prediction of intriguing physics forbidden in ordinary Hermitian models, such as asymmetric scattering and unidirectional or bidirectional invisibility of the potential \cite{r10,r11,r12,r13,r14,r15,r16,r17}.\\ 
In a seminal paper, S.A.R. Horsley 
and collaborators have introduced the class of Kramers-Kronig complex potentials \cite{r18},  in which the spatial profiles of the real and imaginary parts of the potentials are related one another by a Hilbert transform. The properties of such newly discovered potentials, i.e. unidirectional or bidirectional transparency, invisibility and some sublets related to the slow decay of the potentials,  have been theoretically  investigated in a couple of subsequent works \cite{r19,r20,r21,r22,r23,r24,r25,r26,r26bis}, with recent attempts to experimentally realize such a kind of complex potentials \cite{r27,r27bis}. \\
In all previous studies, wave propagation was formulated in the framework of the Helmholtz or the stationary Schr\"odinger equations, which are suited to describe scattering phenomena of waves in continuous systems. However, in several physical contexts, such as in quantum or classical transport on a lattice \cite{r28,r29,r30,r31,r32} or in quantum mechanical models with discretized space \cite{r33,r34,r35}, wave transport is better described by the discrete version of the Schr\"odinger equation. Like for the continuous Schr\"odinger equation, reflectionless potentials can be constructed for the discrete Schr\"odinger equation as well \cite{r36,r37,r38,r39,r40}, for example using  the discrete version of supersymmetry or the Darboux transformation. Such previous works could not find any substantial different behavior of supersymmetric-synthesized  scatteringless potentials when space is discretized. However, the continuous and discrete versions of the Schr\"odinger equation may show distinctly different behaviors, which arise mainly for the limited energy band imposed by the lattice as opposed to parabolic dispersion curve in the continuous limit. \par 
 In this work we consider Kramers-Kronig potentials for the discrete version of the Schr\"odinger equation and highlight some very distinct features of wave scattering on a lattice as compared to the continuous Schr\"odinger equation. While in the latter case a Kramers-Kronig potential is unidirectionally or bidirectionally reflectionless, a stationary Kramers-Kronig potential on a lattice is reflective, i.e. discretization of space breaks the reflectionless property of the Kramers-Kronig potentials. However, we show that a class of slowly drifting Kramers-Kronig potentials on a lattice can become invisible. Our results disclose a very distinct scattering behavior of Kramers-Kronig potentials in continuous and discrete  Schr\"odinger equation models, and are expected to stimulate further theoretical and experimental investigations of such an important class of recently discovered complex potentials.

\section{Wave Reflection from a moving potential on a lattice}
\subsection{Drifting potential on a lattice: Basic equations}

We consider wave reflection from a drifting potential on a one-dimensional lattice, which is described by the discrete Schr\"odinger equation for the wave amplitude $\psi(x,t)$ \cite{r29,r30,r41,r42}
\begin{equation}
i \frac{\partial \psi}{\partial t}= -2 \kappa \cos (a \hat{p}_x) \psi+ V(x+vt) \psi
\end{equation}
i.e.
\begin{equation}
i \frac{\partial \psi}{\partial t}= -\kappa [\psi(x+a,t)+\psi(x-a,t) ]+ V(x+vt) \psi,
\end{equation}
where $\hat{T}=-2 \kappa \cos(a \hat{p}_x)=- \kappa [\exp(a \partial_x)+\exp(-a \partial_x)]$ is the kinetic energy operator, $a$ is the lattice period, $x$ is the spatial variable defined on the discrete sites $x=na$ ($n=0, \pm1 , \pm2, ...$), $\hat{p}_x=-i \partial_x$ is the momentum operator, $V(x)$ is the scattering potential and $v$ is the drift velocity. The parameter $\kappa$ entering in the kinetic energy operator is the hopping rate which determines the width of the tight-binding lattice band. The dispersion relation of the lattice band is sinusoidal and given by $E(q)=-2 \kappa \cos(qa)$, where $q$ is the Bloch wave number. The continuous limit is obtained for a small lattice period $a$ after setting $\cos( a \hat{p}_x) \simeq 1-(a^2/2) \hat{p}_x^2$ in Eq.(1). In this limit the discreteness of space is lost and one obtains the continuous Schr\"odinger equation
\begin{equation}
i \frac{\partial \psi}{\partial t}=- \kappa a^2 \frac{\partial^2 \psi}{\partial x^2}-2\kappa \psi+V(x+vt) \psi
\end{equation}
with the usual parabolic dispersion relation $E(q)=-2\kappa+ \kappa a^2q^2$ of the kinetic energy term.\\
The scattering potential $V(x)$ is assumed to vanish as $ x \rightarrow \pm \infty$ sufficiently fast so as the asymptotic solutions to Eqs.(2) and (3) far from the scattering potential are plane waves. To study the scattering problem, it is convenient to write Eq.(2) in the reference frame of the drifting potential via the Galileian transformation
\begin{equation}
X=x+vt \; , \;\; T=t.
\end{equation} 
This yields the transformed equation
\begin{equation}
i \frac{\partial \psi}{\partial T}=- \kappa [ \psi(X+a,T)+\psi(X-a,T)]+V(X) \psi -iv \frac{\partial \psi}{\partial X}
\end{equation}
 which differs from Eq.(2) owing to the drift term on the right hand side of Eq.(5). Note that, after the Galileian transformation (4), the continuous limit of the Schr\"odinger equation [Eq.(3)] takes the form
 \begin{equation}
 i \frac{\partial \psi}{\partial T}=-2 \kappa \psi -\kappa a^2 \frac{\partial^2 \psi}{\partial X^2}+V(X) \psi-i v \frac{\partial \psi}{\partial X}
 \end{equation}
 which again differs from the original equation because of a drift term [the last term on the right hand side of Eq.(6)]. In such a continuous limit, the drift term can be removed via a gauge transformation and the continuous Schr\"odinger equation is thus invariant under a Galileian transformation. In fact, after the gauge transformation $\psi(X,T)=\phi(X,T) \exp(-i \beta T+i \gamma X)$  
with $\gamma=-v/(2 \kappa a^2)$ and $\beta=-v^2/(4\kappa a^2)$, one can readily show that $\phi(X,T)$ satisfies Eq.(6) but without the drift term on the right hand side.  
 This result is basically due to the fact that the continuous Schr\"odinger equation is a non-relativistic wave equation, and it is therefore invariant under a Galileian transformation \cite{r43}. Such an invariance ensures that the scattering properties of the potential $V(x)$ are not changed when it drifts at a uniform speed $v$: in the laboratory reference frame $(x,t)$, the main effect of the moving potential is a Doppler shift of the frequency of the scattered (reflected) wave. However, for the discrete Schr\"odinger equation (5) in the moving reference frame the drift tern can not be removed via a gauge transformation, i.e. the discrete Schr\"odinger equation is not invariant under a Galileian transformation. This result basically stems from the discrete translational symmetry of the lattice, so that in the reference frame $(X,T)$ the scattering potential is at rest however the lattice is drifting in the opposite direction. A major impact of the breakdown of Galileian invariance for the discrete Schr\"odinger equation is that the scattering properties of a potential $V(x)$ on a lattice are modified when the potential drifts, as we are going to show in the following analysis.  
 \subsection{Reflection and transmission coefficients in the moving reference frame}
 Let us first consider the case of a vanishing scattering potential $V(X)=0$. Then in the moving reference frame the scattering solutions to Eq.(5) are plane waves $\psi(X,T) \propto \exp(iqX-iET)$ with Bloch wave number $q$ and energy $E=E(q)$ defined by the dispersion relation
 \begin{equation}
 E(q)=-2 \kappa \cos (qa)+qv
 \end{equation}
 and group velocity
 \begin{equation}
 v_g(q)=\frac{\partial E}{\partial q}=2 a \kappa \sin (qa)+v.
 \end{equation}
 A typical behavior of the dispersion curve $E=E(q)$ is shown in Fig.1 for increasing values of the drift velocity $v$. Note that, in the moving reference frame, the energy dispersion curve  acquires a linear ramp term $qv$ which breaks the periodicity of $E(q)$.\\ 
   \begin{figure}[htbp]
  \includegraphics[width=83mm]{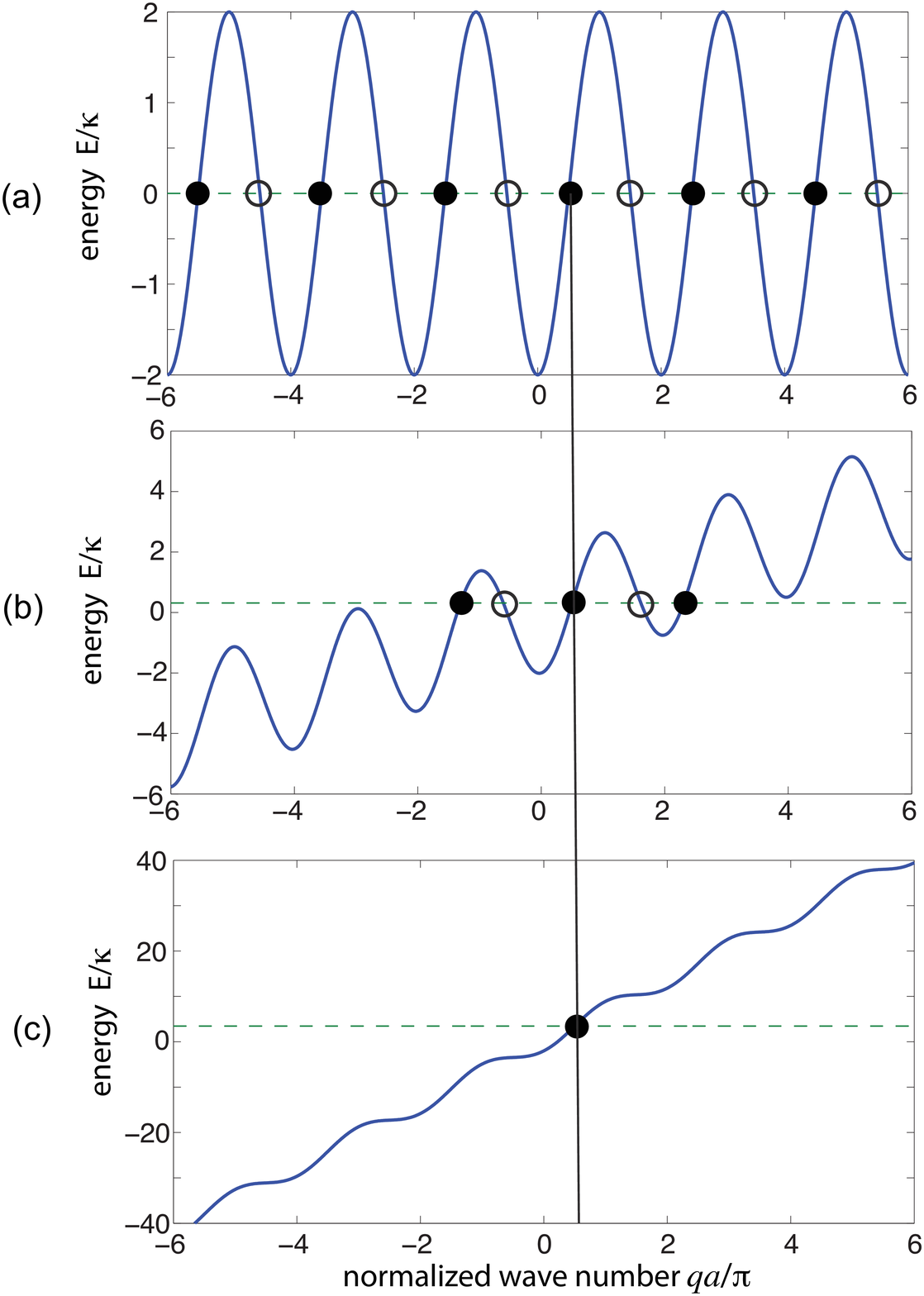}\\
   \caption{(color online) Behavior of the energy dispersion curve $E(q)=-2 \kappa \cos (qa)+qv$ of the lattice in the moving reference frame for increasing values of the drift velocity: (a) $v=0$ (lattice at rest), (b) $v / \kappa=0.2a$, and (c) $v/ \kappa =2.2a$. The horizontal dashed line is the energy $E_0$ of the incident plane with wave number $q_0$ ($q_0a= \pi/2$ in the figure, shown by the vertical solid line). The roots $q_{\beta}$ and $Q_{\alpha}$ of the equation $E(q)=E_0$, corresponding to Bloch waves with positive and negative group velocities, are shown by filled and empty circles, respectively. Note that, for a drift velocity $v$ larger than the critical velocity $v_c= 2 a \kappa$ [as in (c)], the equation $E(q)=E_0$ is satisfied solely for $q=q_0$.}
\end{figure}

 Let us now consider a scattering potential $V(X)$ which vanishes sufficiently fast as $|X| \rightarrow \infty$ so that the scattering solutions to Eq.(5) with energy $E$ are asymptotically plane waves.  To study the scattering problem, for the sake of definiteness we will assume $v>0$  and will  consider a forward-propagating plane wave with Bloch wave number $q_0$ and positive group velocity $v_g(q_0)>0$ (left incidence side), however the analysis can be readily extended to the $v<0$ case or to the right incidence side. Note that the limit of a non-drifting  potential is obtained by letting $v=0$. Since in the moving reference frame $(X,T)$ the scattering potential $V(X)$ is at rest, scattering of a plane wave with defined energy $E_0$ is elastic, i.e. it conserves the energy, and the solution to Eq.(5) corresponding to an incoming plane wave from the left side with wave number $q_0$ is then of the form $\psi(X,T)=f(X) \exp(-i E_0 T)$, where $E_0=E(q_0)$ and $f(X)$ satisfies the stationary differential-difference equation
 \begin{equation}
E_0 f(X)=-\kappa [f(X+a)+f(X-a)]+V(X)f(X)-iv \frac{df}{dX}
 \end{equation}
 with the asymptotic behavior
 \begin{equation}
 f(X)  \sim 
 \left\{
 \begin{array}{ll}
  \exp(i q_0 X)+ \sum_{\alpha} r_{\alpha}(q_0) \exp(iQ_{\alpha} X) & X \rightarrow - \infty \\
  \sum_{\beta} t_{\beta}(q_0) \exp(i q_{\beta} X) & X \rightarrow \infty 
 \end{array}
 \right.
 \end{equation}
 In Eq.(10), the wave numbers $Q_{\alpha}$ and $q_{\beta}$ are defined as the real roots of the equation $-2 \kappa \cos (qa)+vq=E_0$ with $v_g(q_{\beta}) \geq 0$ and $
v_g(Q_{\alpha})<0$; see Fig.1(b). They correspond to the wave numbers of reflected and transmitted plane waves with the same energy $E_0$ than the incident wave, $r_{\alpha}$ and $t_{\beta}$ being the reflection and transmission amplitudes, respectively. Note that, for $\beta=0$, $q_{\beta}=q_0$ is precisely the wave number of the incident wave. The number of the roots $Q_{\alpha}$ and $q_{\beta}$ depends sensitively on the drift velocity $v$, and increases as $v \rightarrow 0$, as schematically shown in Fig.1 \cite{note0}. For a drifting potential with a speed $v$ larger than the critical velocity $v_c \equiv 2 \kappa a$, $\{Q_{\alpha} \}$ is empty, whereas $\{q_{\beta} \}$ is composed solely by the wave number $q_0$ of incident wave [Fig.1(c)]: this means that elastic scattering forbids wave reflection from any potential \cite{note}. Here we focus our analysis to a slowly drifting potential $v< v_c$, for which elastic scattering permits wave reflection. 

\section{Scattering from a Kramers-Kronig potential on a lattice}
Unlike for the continuous Schr\"odinger equation, a Kramers-Kronig potential at rest on a lattice is not reflectionless. The main physical reason of such a result is schematically illustrated  in Fig.2 and can be explained as follows. Let us consider a plane wave with wave number $q_0$, corresponding to a positive group velocity (progressive wave) which is scattered off by a Kramers-Kronig potential $V(x)$ which is holomorphic, for the sake of definiteness, in the upper half complex plane ${\rm Im}(x) \geq 0$. The analyticity of the potential in the half complex plane ensures that its Fourier spectrum $\hat{V}(k)=\int dx V(x) \exp(-ikx)$ vanishes for any $k <0$, i.e. it is composed solely by positive wave numbers, depicted by the solid thin arrows in Fig.2. Therefore, at any scattering order the scattered waves have wave numbers which can not be smaller than $q_0$. In the continuous limit, the Schr\"odinger equation shows a parabolic energy dispersion curve [Fig.2(a)], meaning that all scattered waves have a positive group velocity, i.e. reflection is cancelled. Conversely, for the discrete Schr\"odinger equation [Fig.2(b)] the energy dispersion curve is sinusoidal, so that scattered waves with a wave number larger than $q_0$ may correspond to a negative group velocity, i.e. reflection is allowed.  Figure 3 shows, as an example, reflection of a Gaussian wave packet from a stationary Kramers-Kronig potential on a lattice as obtained by numerical simulations of Eq.(1), for both left and right incidence sides, in the $(x,t)$ laboratory reference frame. The numerical method of integration is described at the end of the section. The scattering potential used in the simulations is given by $V(x)=V_0(x) \exp (i \Omega x)$ with $V_0(x)=V_0 / (x/a+i \alpha)^2$ and with parameter values $V_0 / \kappa=i$, $\Omega=10/a$ and $\alpha=0.3$. Note that for both left and right incidence sides the potential is not reflectionless.\\
 The main result of the present work that  we are going to demonstrate is that, under certain conditions, a class of Kramers-Kronig potentials which are reflective at rest become reflectionless (and even invisible) when drifting on the lattice. As discussed in the previous section, such a result stems from the fact that the discrete Schr\"odinger equation is not invariant under a Galileian transformation, so that the scattering properties of a potential on a lattice change with the drift velocity of the potential. Precisely, we can prove the following general theorem, which states a sufficient condition for a slowly-drifting Kramers-Kronig potential to be invisible:\\
\begin{figure}[htbp]
  \includegraphics[width=83mm]{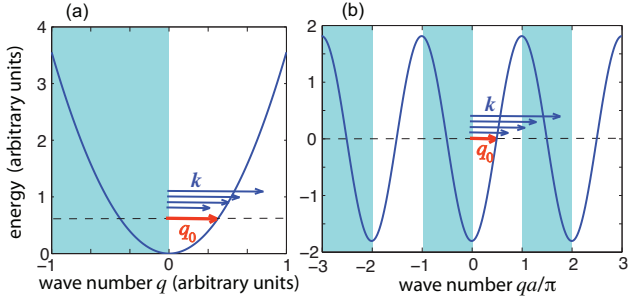}\\
   \caption{(color online) Schematic of the scattering process of a progressive plane wave with wave number $q_0$ from a Kramers-Kronig potential at rest composed by positive wave number components for (a)  the continuous Schr\"odinger equation with parabolic dispersion curve, and (b) the  discrete Schr\"odinger equation with sinusoidal dispersion curve. The thin solid arrows depict the wave number components $k$ of the scattering potential. The shaded areas correspond to waves with negative group velocity, i.e. to reflected waves. The horizontal dashed lines define the energy $E_0$ of the incoming wave. Scattered waves are obtained by mixing the wave numbers of incoming wave and of scattering potential. For elastic scattering energy must be conserved. In (a) reflection is forbidden because there are not excited scattered waves with negative group velocities, whereas in (b) reflection is permitted.}
\end{figure}
\begin{figure}[htbp]
  \includegraphics[width=83mm]{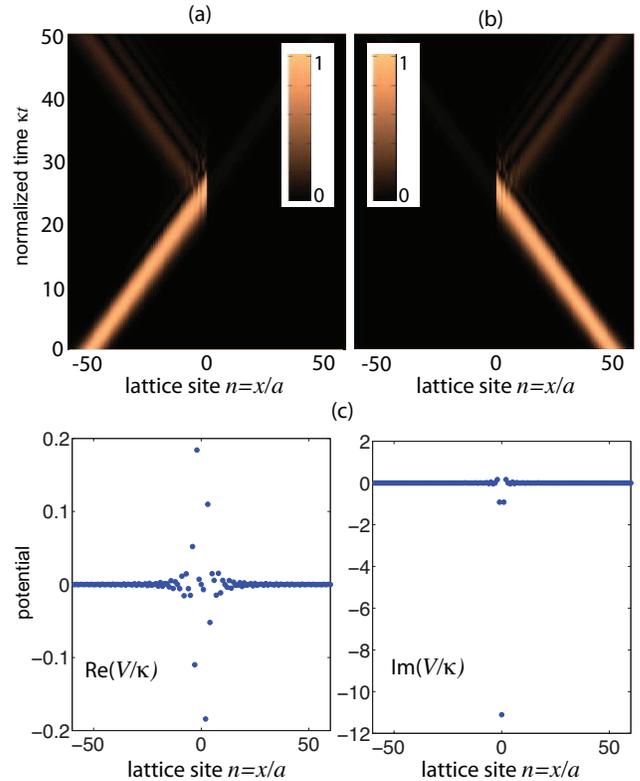}\\
   \caption{(color online) Scattering of a Gaussian wave packet on a lattice from a Kramers-Kronig potential at rest for (a) left, and (b) right incidence sides. The two panels show snapshots of $|\psi(x,t)|$ on a pseudo color map. The scattering potential, defined on the lattice sites $n=x/a$, is given by $V(x)=i \kappa  \exp(i 10 x/a) /(x/a+0.3 i)$ and is shown in panel (c) (real and imaginary parts of $V / \kappa$). The initial condition is the Gaussian wave packet $\psi(x,t=0) \propto \exp[-(x-d)^2/w^2]+i q_0 x]$ with $q_0= \pi/(2a)$, $d=-50a$ and $w=5a$ in (a), and $q_0= -\pi/(2a)$, $d=50a$ and $w=5a$ in (b).}
\end{figure}
\begin{figure}[htbp]
  \includegraphics[width=83mm]{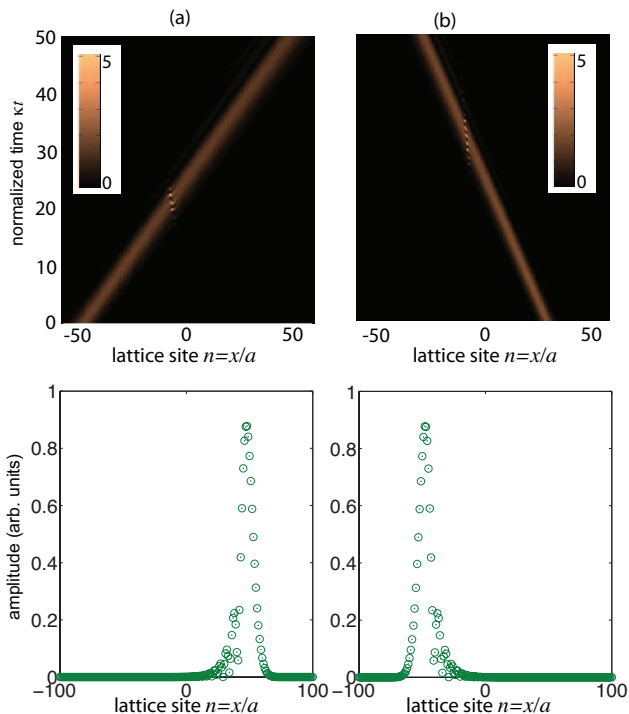}\\
   \caption{(color online) Scattering of a Gaussian wave packet from a drifting Kramers-Kronig potential for (a) left, and (b) right incidence sides. The upper panels show snapshots of $|\psi(x,t)|$ on a pseudo color map. The bottom panels show the detailed distributions of the amplitudes $|\psi(x,t)|$ at final time $t=50 / \kappa$, after the scattering process (open circles), and compare them with those that one would observe in the absence of the scattering potential (points). The potential $V(x)$ is the same as in Fig.3. The drift velocity is $v=0.4 \kappa a$. Initial conditions are as in Fig.3.}
\end{figure}
\begin{figure}[htbp]
  \includegraphics[width=83mm]{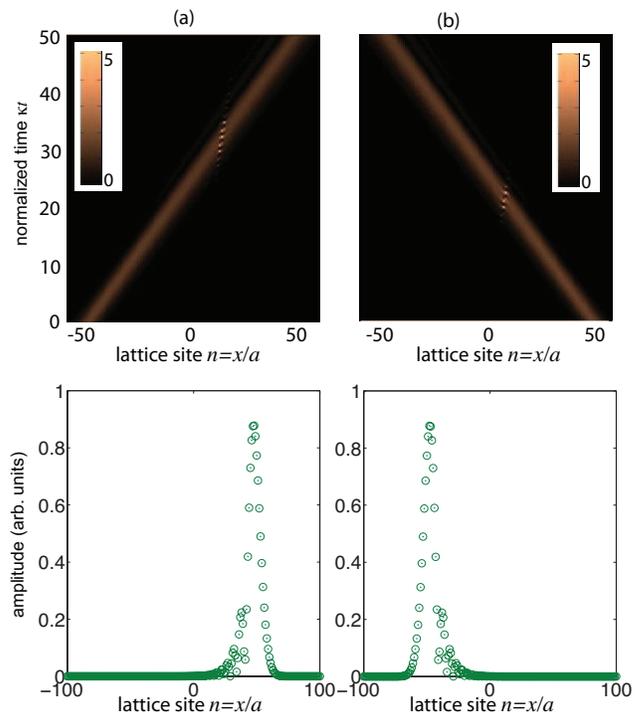}\\
   \caption{(color online) Same as Fig.4, but for a reversed drift velocity  $v=-0.4 \kappa a$.}
\end{figure}
{\em Let $V(X)$ be a Kramers-Kronig potential of the form $V(X)=V_0(X) \exp(i \Omega X)$, with $V_0(X)$ holomorphic in the ${\rm Im}(X) \geq 0$ half complex plane, drifting on a lattice with a speed $v$ smaller than the critical speed $v_c= 2 \kappa a$. Then for $\Omega \geq 4 \kappa /v$ the potential is bidirectionally invisible \cite{note2}.}\\
To prove the theorem, we follow a procedure similar to the one illustrated in Refs.\cite{r20,r25} and based on the complex spatial displacement method. Let us consider, for the sake of definiteness, a progressive wave incident from the left side, so that the asymptotic form of the scattered solution is given by Eq.(10).  Since $V(X)$ is holomorphic in the half complex plane ${\rm Im}(X) \geq 0$, the solution to Eq.(9) can be analytically prolonged from the real $X$ axis into such a half plane. In particular, let us indicate by $f(\xi, \delta)=f(X=\xi+i \delta)$ the solution to Eq.(9) on the horizontal line $\Gamma$ defined by the parametric equation $X=\xi+i \delta$, with fixed $\delta>0$ and $-\infty < \xi < \infty$, and with the asymptotic form defined by Eq.(10) as $\delta \rightarrow 0^+$. The main idea of the complex spatial displacement method is to find suitable connection relations between reflection and transmission amplitudes of scattered waves on the real $X$ axis, i.e. for $\delta=0$, and on the line $\Gamma$, i.e. for $\delta>0$. Since for $\delta \rightarrow \infty$ the scattering potential $V(X)$ vanishes, the reflection and transmission amplitudes of scattered waves on the line $\Gamma$ can be readily determined by perturbative methods for $\delta$ large. The connection formulas can then be used to compute reflection and transmission amplitudes of the original problem, i.e. on the real axis $\delta=0$. For $\delta >0$, the asymptotic form of $f(\xi, \delta)$ as $\xi \rightarrow \pm \infty$ is given by
\begin{equation}
 f(\xi, \delta)  \sim 
 \left\{
 \begin{array}{r}
  A(\delta) \left[ \exp(i q_0 \xi)+ \sum_{\alpha} r_{\alpha}(q_0,\delta ) \exp(iQ_{\alpha} \xi) \right]  \\
    \xi \rightarrow - \infty \\
  A(\delta) \sum_{\beta} t_{\beta}(q_0,\delta) \exp(i q_{\beta} \xi) \\
   \xi \rightarrow \infty 
 \end{array}
 \right.
 \end{equation}
 with $A(\delta=0^+)=1$. As in Eq.(10), in Eq.(11) $r_{\alpha}(q_0,\delta )$ and $t_{\beta}(q_0,\delta)$ are the reflection and transmission amplitudes of scattered waves on the line $\Gamma$, which reduce to  $r_{\alpha}(q_0)$ and $t_{\beta}(q_0)$ entering in Eq.(10) in the $\delta \rightarrow 0^+$ limit. Since $f(X)$ is an analytic function of $X=\xi+i \delta$, the following relation holds
 \begin{equation}
 \frac{\partial f}{\partial \delta}=i \frac{\partial f}{\partial \xi}.
 \end{equation}
Using Eqs.(11) and (12), it readily follows that $A(\delta)= \exp(-q_0 \delta)$. Moreover, the following connection formulas for reflection and transmission amplitudes on the real $X$ axis and on the line $\Gamma$ are found
\begin{eqnarray}
r_{\alpha}(q_0,0) & = & r_{\alpha}(q_0,\delta) \exp [-(q_0-Q_{\alpha}) \delta] \\
t_{\beta}(q_0,0) & = & t_{\beta}(q_0,\delta) \exp [-(q_0-q_{\beta}) \delta] .
\end{eqnarray}
 For $\delta \rightarrow \infty$, the potential $V(X= \xi+i \delta)$ vanishes and the order of magnitude of $r_{\alpha}(q_0,\delta)$, 
$t_{\beta}(q_0,\delta)$ can be estimated by first-order Born approximation \cite{r25}. As shown in the Appendix, $r_{\alpha}(q_0,\delta) \rightarrow 0$ and $t_{\beta}(q_{0},\delta) \rightarrow 0$ ($\beta \neq 0$) at least like $\sim \exp(-\delta \Omega)$, whereas $t_0(q_0,\delta) \rightarrow 1$. Provided that the condition $\Omega \geq  4 \kappa / v$ is met, $\Omega$ is always larger than any difference $|q_0-q_{\beta}|$ and $|q_0-Q_{\alpha}|$. Therefore from Eqs.(13) and (14) one obtains $r_{\alpha}(q_0,0)=0$, $t_{\beta}(q_0,0)=\delta_{\beta,0}$, which means that the scattering potential $V(X)$ is invisible for left incidence side. A similar proof can be done assuming a wave incident from the right side, i.e. the potential $V(X)$ is bidirectionally invisible. \par

We checked the bidirectional invisibility of moving Kramers-Kronig potentials by direct numerical simulations of the discrete Schr\"odinger equation (1) in the laboratory reference frame $(x,t)$. By letting $x=na$ and $c_n(t)=\psi(x=na,t)$, the differential-difference equation (2) is equivalent to the following set of linear coupled equations for the complex amplitudes $c_n(t)$ on the lattice
\begin{equation}
i \frac{dc_n}{dt}=-\kappa(c_{n+1}+c_{n-1})+V(na+vt) c_n
\end{equation}
with time-dependent coefficients. The coupled equations (15) are numerically solved using an accurate variable-step fourth-order Runge-Kutta method assuming open boundary conditions. The lattice size, i.e. number of lattice sites, has been set large enough (typically $-100 \leq n \leq 100$) to avoid edge effects.
As an example, Figs.4 and 5 show numerical results of bidirectional invisibility for a propagating Gaussian wave packet scattered off by the same Kramers-Kronig potential as in Fig.3, but when the potential drifts on the lattice with a velocity $v= \pm 0.4a \kappa$. The figures depict the temporal evolution of the wave packet amplitude $|\psi(x,t)|$, for either left and right incidence sides, and compare the wave packet distributions after the scattering process with the ones observed in the absence of the scattering potential. The coincidence of the distributions is the clear signature that the drifting Kramers-Kronig potential is invisible, while it is reflective at rest.  

  \section{Conclusions and discussion}
  Wave scattering from complex potentials in the framework of non-Hermitian extensions of the Schr\"odinger or Helmholtz equations has received a great and increasing interest in the past few years, with the discovery of intriguing physics forbidden in ordinary Hermitian scattering problems, such as asymmetric reflection and unidirectional or bidirectional invisibility of the potential. An important class of complex potentials, which do no reflect waves from one or both incidence sides, in provided by so-called Kramers-Kronig potentials \cite{r18}, in which the real and imaginary spatial profiles of the potential are related one to another by a Hilbert transform. Most of recent studies focused on wave scattering from non-Hermitian potentials in continuous wave equations, however in several physical systems wave transport is better described by discrete wave equations. A paradigmatic equation describing discrete wave transport is provided by the discrete Schr\"odinger equation, which is encountered in models of quantum or classical transport on a lattice or in quantum mechanical models with discretized space. In this work we have shown that discretization of space and breaking of the continuous translational spatial invariance deeply change the scattering properties of Kramers-Kronig potentials on a lattice. In particular, the physical mechanism that prevents wave reflection of a Kramers-Kronig potential in the continuous Schr\"odinger equation breaks down when scattering occurs on a lattice with discrete translational invariance. Therefore, a Kramers-Kronig potential on a lattice is rather generally a reflective potential. However, we have shown that if the potential slowly drifts on the lattice, under certain conditions it can become bidirectionally invisible. Our study sheds new light into the important and broad field of wave scattering in non-Hermitian physical models and highlights important distinctive features of wave scattering in discrete versus continuous wave equations. In particular, we revealed that breakdown of Galileian invariance in discrete wave equations can enable a reflective potential to become reflectionless when drifting on the lattice.\\ 
  Physically, Kramers-Kronig potentials on a lattice and their reflection properties could be implemented in optics using arrays of evanescently-coupled optical waveguides or chains of microring resonators with tailored gain and loss profiles \cite{r32,refer1}. For example, it is known that spatial light propagation along the longitudinal $z$ axis in a lattice of coupled dielectric optical waveguides is described by coupled-mode equations analogous to Eq.(15), in which time $t$ is replaced by the spatial coordinate $z$ and $n$ is the waveguide number \cite{refer1}. The real and imaginary parts of the potential $V$ can be tailored by controlling, along the propagation distance $z$, the effective mode index of the waveguides, i.e. propagation constant offset of the waveguide mode and optical amplification/attenuation. If optical amplification (gain) is not available, one can resort to a purely dissipative (lossy) structure \cite{refer3}. For example, using waveguide arrays written in a glass with the femtosecond laser writing technique \cite{refer4,refer5,refer6,refer7} an effective propagation constant mismatch can be introduced by varying the writing speed of the focused laser beam in the glass \cite{refer4,refer7}, whereas selective optical losses can be obtained by patterning a selective layer of absorptive material on the top of the array or by suitable bending of waveguides \cite{refer6}. Static and moving potentials are readily obtained by manufacturing different waveguide arrays with straight or transversely tilted perturbation $V$ of the effective mode index. Excitation of the array by a tilted Gaussian beam and monitoring its propagation along the $z$ axis using fluorescence imaging methods \cite{refer1,refer4,refer5} enables one to visualize the wave packet dynamics in the different regimes.\\  
 It is envisaged that our results could stimulate further theoretical and experimental investigations on discrete wave transport and scattering by non-Hermitian potentials. Optical waveguide arrays could provide an experimentally accessible laboratory tool for the observation of the scattering properties of spatial Kramers-Kronig potentials on a lattice. On the theoretical side, the analysis could be extended to a two-dimensional lattice, in which the band structure and transport are known to be deeply modified by synthetic gauge fields. In principle, in a two-dimensional lattice the scattering properties of non-Hermitian potentials could be controlled by synthetic gauge fields, which is not feasible in continuous wave scattering settings. The interplay of non-Hermitian potential scattering and gauge fields could be a subject of future research. 
  
  \appendix
\section{Reflection and Transmission amplitudes of the spatially displaced potential}
Let us indicate by $G(\xi)$ the potential $V(X)=V_0(X) \exp(i \Omega X)$ on the line $\Gamma$ of the upper complex plane, i.e. for $X= \xi+i \delta$, with $\delta>0$ large enough and $-\infty < \xi < \infty$. The displacement $\delta$ on the imaginary axis of the potential pushes all the possible singular behavior of  $V_0(X)$ further down into the
lower complex plane, simultaneously reducing its
magnitude. In particular, since $G(\xi)=G_0(\xi) \exp(i \Omega \xi -\delta \Omega)$ with $G_0(\xi) \equiv V_0(\xi+i \delta)$, for $\delta \rightarrow \infty$ $G(\xi)$ is exponentially small, uniformly over the entire line $\Gamma$, of order smaller than $\sim \exp( -\delta \Omega)$. Therefore, the weak scattering introduced by the vanishingly potential $G(\xi)$ can be computed by first-order (Born) approximation \cite{r25}. The solution to Eq.(9) on the line $\Gamma$, corresponding to an incident plane wave from the left side of wave number $q_0$, is thus given by
\begin{equation}
f(\xi,\delta) = A(\delta) [ f^{(0)}(\xi)+\phi (\xi)]
\end{equation}
where $f^{(0)}(\xi)= \exp(iq_0 \xi)$ is the unperturbed incident plane wave of amplitude $A(\delta)$ and $\phi(\xi)$ is a small correction introduced by the weak scattering potential $G(\xi)$. At first-order (Born) approximation, $\phi(\xi)$ is the solution of the forced linear equation
\begin{eqnarray}
E_0 \phi(\xi)+\kappa [\phi(\xi+a)+\phi(\xi-a)]+iv \frac{d\phi}{d\xi} & = & \nonumber \\
 G_0(\xi) \exp(-\delta \Omega) \exp(i \Omega \xi+i q_0 \xi).
\end{eqnarray}
The solution to Eq.(A2) is formally given by 
\begin{equation}
\phi(\xi)=\frac{1}{2 \pi} \left(  \int_{-\infty}^{\infty} dk \frac{\hat{G}_0(k-q_0-\Omega) \exp(i k \xi )}{E_0+2 \kappa \cos (ka)-vk} \right) \exp(-\delta \Omega)
\end{equation}
where $\hat{G}_0(k)=\int d\xi G_0(\xi) \exp(i k \xi)$ is the Fourier transform of the potential $G_0(\xi)$. Note that, since $G_0(\xi)=V_0(\xi+i \delta)$ and $V_0(X)$ is holomorphic for ${\rm Im}(X) \geq 0$, $\hat{G}_0(k)$ vanishes for $k<0$, so that the integral on the right hand side of Eq.(A3) is actually extended from $k=q_0+ \Omega$ to $k= \infty$. In such a range, the function under the sign of the integral is not singular, since its poles $q_{\beta}$ and $Q_{\alpha}$ lie in the range $k<\Omega+q_0$. For $\delta \rightarrow \infty$, Eq.(A3) thus shows that $\phi(\xi)$ is exponentially vanishing, at least like $\sim \exp(-\delta \Omega)$. A comparison of Eqs.(11) and (A1) indicates that $r_{\alpha}(q_0, \delta)$ and $t_{\beta}(q_0, \delta)$ ($\beta \neq 0$) are vanishing at least like $\sim \exp(-\delta \Omega)$ as $ \delta \rightarrow \infty$, whereas $t_0(q_0, \delta) \rightarrow 1$.  


\begin{thebibliography}{31}
\bibitem{note0}
For a potential at rest, i.e. for $v \rightarrow 0$, the roots of the equation $E(q_0)=E_0$ are infinite and given $q_{\beta}=q_0+2 \beta \pi/a$ and $Q_{\alpha}=-q_0+2 \alpha \pi /a$ ($\alpha, \beta=0 \pm 1, \pm 2, ...$); see Fig.1(a). However, since on the lattice space $x$ is discretized by the lattice period $a$, such waves are degenerate and there is actually only one transmitted wave with wave number $q_0$ and one reflected wave with wave number $Q_0=-q_0$. The degeneracy is generally lift when $v$ is nonvanishing, as shown in Fig.1(b).
\end{thebibliography}

\end{document}